\documentclass[a4paper,12pt]{article}
\pdfoutput=1
\usepackage{epsfig}
\usepackage{cite}
\usepackage{amssymb}
\usepackage{amsfonts}
\usepackage{amsmath}
\usepackage{euscript}
\usepackage{verbatim}
\usepackage{latexsym}
\usepackage{graphicx}

\newskip\humongous \humongous=0pt plus 1000pt minus 1000pt

\newif\ifdtup

\catcode`\@=11

\@addtoreset{equation}{section}

\def\@normalsize{\@setsize\normalsize{15pt}\xiipt\@xiipt
\abovedisplayskip 14pt plus3pt minus3pt%
\belowdisplayskip \abovedisplayskip
\abovedisplayshortskip \z@ plus3pt%
\belowdisplayshortskip 7pt plus3.5pt minus0pt}

\def\small{\@setsize\small{13.6pt}\xipt\@xipt
\abovedisplayskip 13pt plus3pt minus3pt%
\belowdisplayskip \abovedisplayskip
\abovedisplayshortskip \z@ plus3pt%
\belowdisplayshortskip 7pt plus3.5pt minus0pt
\def\@listi{\parsep 4.5pt plus 2pt minus 1pt
     \itemsep \parsep
     \topsep 9pt plus 3pt minus 3pt}}

\relax

\catcode`@=12

\catcode`\@=11

\def\section{\@startsection{section}{1}{\z@}{3.5ex plus 1ex minus
   .2ex}{2.3ex plus .2ex}{\large\bf}}

\def\SymBoxes#1#2#3#4{\newdimen\un@t \un@t#3%
\raisebox{#1}{\rule{#2\un@t}{#4}\hskip-#2\un@t
\@tempdimb\un@t \advance\@tempdimb by-#4\@tempcntb#2\relax%
\@whilenum{\@tempcntb>0}\do{
\rule{#4}{\un@t}\hskip\@tempdimb \advance\@tempcntb by\m@ne}%
\hskip-#2\un@t \rule[\un@t]{#2\un@t}{#4}%
\rule[\un@t]{#4}{#4}\hskip-#4
\rule{#4}{\un@t}}\hskip-#4}                

\begin{document}

\newcommand{\be}{\begin{equation}}
\newcommand{\ee}{\end{equation}}
\newcommand{\bea}{\begin{eqnarray}}
\newcommand{\eea}{\end{eqnarray}}
\newcommand{\beas}{\begin{eqnarray*}}
\newcommand{\eeas}{\end{eqnarray*}}
\newcommand{\defi}{\stackrel{\rm def}{=}}
\newcommand{\non}{\nonumber}
\newcommand{\bquo}{\begin{quote}}
\newcommand{\enqu}{\end{quote}}
\renewcommand{\(}{\begin{equation}}
\renewcommand{\)}{\end{equation}}
\def \eqn#1#2{\begin{equation}#2\label{#1}\end{equation}}
\def\IZ{{\mathbb Z}}
\def\IR{{\mathbb R}}
\def\IC{{\mathbb C}}
\def\IQ{{\mathbb Q}}
\def\de{\partial}
\def\Tr{ \hbox{\rm Tr}}
\def\H{ \hbox{\rm H}}
\def\HE{ \hbox{$\rm H^{even}$}}
\def\HO{ \hbox{$\rm H^{odd}$}}
\def\K{ \hbox{\rm K}}
\def\Im{ \hbox{\rm Im}}
\def\Ker{ \hbox{\rm Ker}}
\def\const{\hbox {\rm const.}}
\def\o{\over}
\def\im{\hbox{\rm Im}}
\def\re{\hbox{\rm Re}}
\def\bra{\langle}\def\ket{\rangle}
\def\Arg{\hbox {\rm Arg}}
\def\Re{\hbox {\rm Re}}
\def\Im{\hbox {\rm Im}}
\def\exo{\hbox {\rm exp}}
\def\diag{\hbox{\rm diag}}
\def\longvert{{\rule[-2mm]{0.1mm}{7mm}}\,}
\def\a{\alpha}
\def\dag{{}^{\dagger}}
\def\tq{{\widetilde q}}
\def\p{{}^{\prime}}
\def\W{W}
\def\N{{\cal N}}
\def\hsp{,\hspace{.7cm}}

\def\br{\nonumber\\}
\def\IZ{{\mathbb Z}}
\def\IR{{\mathbb R}}
\def\IC{{\mathbb C}}
\def\IQ{{\mathbb Q}}
\def\IP{{\mathbb P}}
\def \eqn#1#2{\begin{equation}#2\label{#1}\end{equation}}

\newcommand{\C}{\ensuremath{\mathbb C}}
\newcommand{\Z}{\ensuremath{\mathbb Z}}
\newcommand{\R}{\ensuremath{\mathbb R}}
\newcommand{\rp}{\ensuremath{\mathbb {RP}}}
\newcommand{\cp}{\ensuremath{\mathbb {CP}}}
\newcommand{\vac}{\ensuremath{|0\rangle}}
\newcommand{\vact}{\ensuremath{|00\rangle}}
\newcommand{\oc}{\ensuremath{\overline{c}}}
\newcommand{\Pf}{\mathrm{Pf}}
\newcommand{\tr}{\mathrm{tr}}
\newcommand{\dt}{\mathrm{det}}
\begin{titlepage}
\begin{flushright}
SISSA 60/2011/EP
\end{flushright}
\bigskip
\def\thefootnote{\fnsymbol{footnote}}

\begin{center}
{\Large
{\bf
Dynamical completions of \\
\vspace{0.2in}
generalized O'Raifeartaigh models
}
}
\end{center}

\bigskip
\begin{center}
{\large
Matteo Bertolini$^{a,b}$,
Lorenzo Di Pietro$^a$ and
Flavio Porri$^a$}

\end{center}

\renewcommand{\thefootnote}{\arabic{footnote}}

\begin{center}
$^a$ {SISSA and INFN - Sezione di Trieste\\
Via Bonomea 265; I 34136 Trieste, Italy\\}
\vskip 5pt
$^b$ {International Centre for Theoretical Physics (ICTP)\\
Strada Costiera 11; I 34014 Trieste, Italy \\}
\vskip 5pt
{\texttt{bertmat,dipietro,fporri @sissa.it}}

\end{center}

\noindent
\begin{center} {\bf Abstract} \end{center}
We present gauge theory completions of  Wess-Zumino models admitting supersymmetry breaking vacua with spontaneously broken R-symmetry. Our models are simple deformations
of generalized ITIY models, a supersymmetric theory with gauge group Sp(N), N+1 flavors plus singlets, with a modified tree level superpotential which explicitly breaks 
(part of) the global symmetry. Depending on the nature of the deformation, we obtain effective O'Raifeartaigh-like models 
whose pseudomoduli space is locally stable in a neighborhood of the origin of field space, or in a region not including it. Hence, 
once embedded in direct gauge mediation scenarios, our models can give low energy spectra with either suppressed or
unsuppressed gaugino mass.
\vspace{1.6 cm}
\vfill

\end{titlepage}

\setcounter{footnote}{0}

\section{Introduction}
\label{intro}

A large class (but not all) of dynamical supersymmetry breaking (DSB) models, either with stable or metastable vacua, can be described at low energy by effective Wess-Zumino
models where supersymmetry is broken at tree level, the original O'Raifeartaigh model being the prototype such example. For this reason, in the last few years a 
renewed attention has been devoted to study general properties of O'Raifeartaigh-like models.

According to the Nelson-Seiberg criterion \cite{Nelson:1993nf} a necessary condition for
supersymmetry breaking is the existence of a R-symmetry in the Lagrangian, if the superpotential is generic. Indeed, most O'Raifeartaigh-like models typically have a
$U(1)_{R}$ symmetry. This may pose a phenomenological problem since gaugino (Majorana) mass terms break the R-symmetry. Hence, one would like also the R-symmetry 
to be broken in the vacuum, in viable supersymmetric extensions of the Standard Model. The original O'Raifeartaigh (O'R) model, as well as many of its generalizations,  
does not have this property: while there always exists a pseudomoduli space at the classical level \cite{Ray:2006wk}, after quantum corrections are taken into account 
the supersymmetry breaking vacuum is stabilized at the origin of field space where the R-symmetry is preserved. The situation does not automatically improve if one 
accepts we live in a metastable vacuum. In these situations, one does not need an exact R-symmetry to be present. However, it turns out that models of this guise 
often possess an approximate R-symmetry which is unbroken in the metastable vacuum. For instance, this is the case for the ISS model  \cite{Intriligator:2006dd}.

It is then a natural question to ask under which general conditions an O'R model can have supersymmetry breaking vacua where also the R-symmetry
is spontaneously broken. This was answered in \cite{Shih:2007av} where it was shown that a necessary condition for this to happen is to have in the theory 
chiral superfields with R-charges different from 0 or 2 (this being instead the case for e.g. the original O'Raifeartaigh
model)\footnote{Strictly speaking, the theorem holds for models with a single pseudomodulus. In principle, allowing for a larger pseudomoduli space one 
could circumvent this restriction, e.g. a pseudomodulus could be responsible for supersymmetry breaking and get stabilized at the origin, and a second one 
could instead break the R-symmetry.  This possibility was recently noticed in  \cite{Evans:2011pz,Shadmi:2011mt}, but we are not aware of any explicit model 
in the literature which actually realizes such scenario.}. In these models quantum corrections can stabilize the pseudomodulus at a non-vanishing VEV, hence 
breaking the R-symmetry spontaneously.

The very reason to ask for vacua with broken R-symmetry is to allow for gaugino mass terms. However, when looking for viable extension of the Standard Model, 
this is not enough to ensure gaugino to sfermion mass ratios of order one. The reason can be summarized as follows \cite{Komargodski:2009jf}. A generic 
O'Raifeartaigh model can always be put in the form
\be
\label{wgenor}
W = f X + \frac 12 \left( M^{ij} + X N^{ij} \right)  \Phi_i \Phi_j + \dots ~,
\ee
where $i=1,\dots,r$, and  ... denote possible cubic terms involving $\Phi_i$ superfields, only \cite{Ray:2006wk}. Supersymmetry is spontaneously broken at tree level
along the one-dimensional pseudomoduli space parametrized by the field $X$ with $\Phi_i=0$. Notice that $X$ has R-charge equal to 2. Using R-symmetry arguments 
one can show \cite{Cheung:2007es} that the following formula holds
 \be
 \label{nN}
 \det \left( M + XN\right) = X^n G(M,N)
~,
\ee
where $n \geq 0$ is an integer and $G(M,N)$ is some function of the couplings.

There is a sharp distinction in the nature of the pseudomoduli space between models with $n=0$ and $n>0$. In the former case $\det M \not =0$ and 
$\det N=0$, and the pseudomoduli space is classically locally stable in a neighborhood of the origin (at least). Models of this kind have been dubbed type 
I in \cite{Cheung:2007es}. When $n>0$, $\det M =0$ and the pseudomoduli space is locally stable only for $|X|>X_{min}$, for some non-vanishing $X_{min}$. 
Models of this kind can be further divided in two subclasses, depending on whether $\det N \not =0$ (type II) or $\det N=0$ (type III). In the latter case 
the stability region is bounded from above, too, i.e. $X_{min} < |X| < X_{max}$. In the following we will adhere to this terminology.

Phenomenologically, the distinction between models with $n=0$ and $n>0$ is even  sharper: when models
as (\ref{wgenor}) emerge as low energy descriptions of some hidden sector and (a subset of) the fields $\Phi_i$ play the role of messengers in a gauge mediation
scenario, the leading order expression for gaugino masses is \cite{Giudice:1997ni}
\be
\label{mgt}
m_{\tilde g} \sim F_X^\dagger \frac{\partial}{\partial X} \log \det \left(M+ NX\right)~.
\ee
From eq.~(\ref{nN}) it follows that models with $n=0$ have suppressed gaugino to sfermion mass ratios, since the contribution (\ref{mgt}) vanishes. 
On the contrary, having $n>0$ is a necessary condition for unsuppressed gaugino masses. 

In this sense, O'R-like models with $n>0$ are possibly the most interesting ones, from a phenomenological view point, and it would be desirable 
to find realizations for such theories as effective DSB models 
 (as well as a guiding principle towards their construction). In this paper we accomplish this task.

\subsection{Summary of results}

The models we propose are simple deformations of DSB theories with quantum deformed moduli space, the prototype example being the ITIY model 
\cite{Intriligator:1996pu,Izawa:1996pk}, a supersymmetric $SU(2)$ gauge theory with 2 flavors plus singlets. Our strategy is similar to the one recently 
pursued in \cite{Goodman:2011jg} (see also \cite{Shadmi:2011uy,Lin:2011vd}), where a completion of a model with $n=0$ and $r=3$ has been proposed. Here we show that suitable deformations of ITIY models 
with gauge group $Sp(N)$ can provide completions of a large class of models with either $n=0$ {\it or} $n>0$, and arbitrary $r$. 
Hence, while providing DSB models which might serve as possible 
hidden sectors in a modular gauge mediation scenario, our models, once embedded into direct gauge mediation, can give, in principle, a soft spectrum with either 
suppressed or unsuppressed gaugino mass\footnote{Our deformations are similar in form to those considered in \cite{Katz:2007gv}. In that work the 
focus was on the existence of ISS-like metastable vacua in SQCD with quantum deformed 
moduli space (which that analysis could not confirm). In order to recover an ISS-like theory, the authors of \cite{Katz:2007gv} take a decoupling limit for all singlets 
of the parent (massive) ITIY model. 
In our models singlets are never completely decoupled, and include a classical pseudomodulus. Moreover, no mass term for the electric quarks is added. Therefore, 
our theory is different.}. 

In fact, the problem of getting R-symmetry breaking vacua has been addressed by many authors in the context of ISS-like 
models (see e.g.\cite{Kitano:2006xg,Csaki:2006wi,Abel:2007jx,Haba:2007rj,Giveon:2008wp,Intriligator:2008fe,Essig:2008kz,Amariti:2008uz,Giveon:2009yu,Abel:2009ze,Barnard:2009ir,Maru:2010yx,Curtin:2010ku}). One basic difference with our models that we would like to emphasize, is that in those constructions the R-symmetry of the UV theory is explicitly 
broken by mass terms, 
and an approximate R-symmetry emerges in the low energy effective theory. The latter is then spontaneously or explicitly broken thanks to suitable modifications 
of the original ISS Lagrangian. Here, we start instead from a gauge theory admitting a non anomalous R-symmetry, and this is hence the same R-symmetry enjoyed in 
the IR, which then happens to be spontaneously broken in the full theory. 

The remainder of this paper is organized as follows. In section 2 we briefly review ITIY models with symplectic gauge groups and the mechanism by which supersymmetry 
gets dynamically broken. At low energy, these models reduce to O'R models with all fields having R-charges either 0 or 2, and hence unbroken R-symmetry. In section 3 
we outline the general strategy one should follow in order to get at low energy O'R models with negative R-charges and $n\geq 0$. Basically, this amounts to add tree level superpotential deformations which partially break the global symmetry of the original ITIY model (keeping, still, the UV theory generic 
and renormalizable). To make our discussion concrete, in section 4 we focus on a specific class of deformations and analyze the corresponding theory in full detail, 
showing explicitly how our strategy works. As DSB models, our models are uncalculable, in the sense that there does not exist a region of the parameter space where K\"ahler corrections can be computed exactly, as much as the original ITIY model. However, following \cite{Chacko:1998si}, 
we show there exists a region of the parameter space where uncalculable K\"ahler corrections are suppressed with respect to those coming from the one loop effective 
potential. Remarkably, this region coincides with that for which the lifetime of the supersymmetry breaking vacua is parametrically large, hence making the full 
construction self-consistent. Section 5 contains our conclusions, and an outlook on possible further investigations and potential applications.


\section{Review of generalized ITIY models}

In this section we briefly review the structure of ITIY models with symplectic gauge group, following \cite{Chacko:1998si}. Let us consider a supersymmetric gauge 
theory with gauge group $Sp(N)$, $F=N+1$ fundamental flavors $Q_i$, $i=1,\dots,2F$, and an antisymmetric singlet $S^{ij}$ with tree level superpotential
\be
W_{tree}=\lambda S^{ij}Q_iQ_j~,
\ee
which respects the $SU(2F)$ flavor symmetry.
When $\lambda=0$ the classical moduli space is parametrized by gauge invariant operators $V_{ij}=Q_iQ_j$ subject to the constraint $\Pf V=0$. For $\lambda \neq 0$ the
mesonic flat directions are lifted, and one is left with a moduli space spanned by $S^{ij}$ with $V=0$. The modified constraint due to non perturbative gauge dynamics
\be
\label{qc}
\Pf V=\Lambda^{2F}
\ee
is therefore incompatible with $\lambda\neq0$ and supersymmetry is broken.

If $\lambda\langle S\rangle \ll \Lambda$, quarks are light at tree level and the low energy theory is rewritten in terms of the meson matrix $V$. The independent 
degrees of freedom are determined by solving the quantum constraint (\ref{qc}), and they can be identified with the fluctuations around a point of the quantum 
moduli space. At a generic such point the global $SU(2F)$ symmetry is broken to $SU(2)^F$, but there are submanifolds of enhanced symmetry. Since we want the 
chosen point to be the real (meta)stable minimum once the susy breaking mechanism is taken into account, we choose a point belonging to the compact submanifold 
of maximal symmetry $Sp(F)$ \footnote{We thank Zohar Komargodski for a discussion on this point.}, and solve the constraint in an expansion around it
\be
V = \Lambda(V_0 J + V')\,,\quad S = \frac{1}{\sqrt{2F}}S_0 J + S'~,
\ee
where $J$ is the $Sp(F)$ invariant tensor and $V'$, $S'$ satisfy $\tr[JV']=0=\tr[JS']$. The factor of $\Lambda$ in the definition above is to make the dimension
of $V_0$ and $V'$ fields equal to one. The solution of the quantum constraint for $V_0$ in a small $V'$ expansion is
\be
V_0 = \Lambda \left( 1 - \frac{1}{4F\Lambda^2}\tr\left[JV'JV'\right] + \mathcal{O}\left(\frac{V'^3}{\Lambda^3}\right)\right).
\ee
giving us the following low energy superpotential to quadratic order in $V'$
\be
W_{eff}=fS_0 + h\,S_0 \Tr\left[JV'JV'\right] - \lambda \Lambda\Tr\left[S'V'\right]~,
\ee
where 
\be
\label{fh}
f=\sqrt{2F}\lambda\Lambda^2~~\mbox{and}~~ h = - \frac{1}{2\sqrt{2F}}\lambda ~. 
\ee
Upon the identification $S_0 \equiv X , (V',\,S') \equiv \Phi_i$ we see that we get
an O'R model of the form (\ref{wgenor}), with $\dt M \neq 0$, $Sp(F)\times U(1)_R$ global symmetry and all R-charges equal 0 or 2. The pseudomoduli space is hence
stabilized at the origin of field space by the one loop potential. As discussed in \cite{Chacko:1998si}, these perturbative corrections are dominant with respect to
uncalculable K\"ahler contributions at least near the origin of the pseudomoduli space, making the existence of the supersymmetry breaking minimum (which is a global 
one, in this case) reliable.

\section{Modified ITIY models}

In what follows we want to discuss which modifications one could make on the model described above to obtain more general O'R models at low energy. As already stressed,
our goal is to obtain models with R-charges different from 0 and 2 and, possibly, with both $n=0$ and $n>0$. From general arguments \cite{Cheung:2007es}, we expect the 
supersymmetry breaking minimum to be at most metastable, as lower energy vacua are expected to emerge in these more general models. We will find indeed runaway vacua 
in the effective theories, which, as we will see, may or may not be real runaways in the full theory.

In this section we 
outline the general strategy one should follow in order to accomplish our task. In section 4 we will put our general recipe at work, focusing on some (classes of) 
models and discussing in an explicit example the full dynamics in detail.

As recently discussed in \cite{Goodman:2011jg}, a simple way to obtain fields with R-charge different from 0 and 2 is to break the global symmetry and make the 
original anomaly-free R-symmetry to mix with some broken $U(1)$ generator of the flavor symmetry group. To this end, one can add explicit symmetry breaking terms 
in the superpotential and/or reduce the field content of the theory. Suppose that in a way or another the global symmetry is broken according to the pattern
\be
\label{bfg}
SU(2F)\times U(1)_R \to G \times U(1)_{R'}~,
\ee
where $G$ is a subgroup of the residual $Sp(F)$ around the enhanced symmetry point of the moduli space. Recall that $V'$ is in an irreducible representation of 
$Sp(F)$ which we denote by $\boldsymbol{r}$ and $S'$ is in the conjugate representation $\boldsymbol{\bar{r}} = (\boldsymbol{r}^T)^{-1}$. Such representations 
split in irreducible $G$ representations as
\bea
\label{repvs}
\boldsymbol{r} = \boldsymbol{r_1} \oplus \dots \oplus \boldsymbol{r_k}  ~,~ & V' = (V_1,\dots, V_k)\nonumber \\
\boldsymbol{\bar{r}} = \boldsymbol{\bar{r}_1} \oplus \dots \oplus \boldsymbol{\bar{r}_k} ~,~ & S' = (S_1,\dots, S_k) 
~.
\eea
The two representations are also equivalent, hence each block in $\boldsymbol{r}$ decomposition is equivalent to a certain one in $\boldsymbol{\bar{r}}$ decomposition. 
Since $JV'J$ is in the same $\boldsymbol{\bar{r}}$ representation of $S'$, the upshot is that the $Sp(F)$ invariant quadratic terms in the $V'$ and $S'$ fields are 
rewritten as
\bea\label{SpInv}
\tr\left[S' V'\right]  =  \sum_{I = 1}^k S_I V_I ~,\nonumber \\
 h\,\tr\left[J V' J V'\right]  =  \sum_{I,J=1}^k C^{IJ}V_I V_J~,
\eea
where contractions of the representations are understood, and the matrix $C$ acts by swapping some couples of indices, i.e. it takes the form
\be
C = \left(\begin{array}{c|c}
       C_1 & 0       \\
       \hline
       0      & C_2  \\
       \end{array}\right)~,~
\ee
with
\be
C_1 = \mathrm{diag}(c_1^{(1)},\dots,c_1^{(p)}) ~,~ C_2 = \mathrm{diag}(c_2^{(1)},\dots,c_2^{(q)})\otimes\left(\begin{array}{cc}
											                             											       0 & 1 \\
											     														       1 & 0
                 										   														     \end{array}\right).
\ee
By genericity, also the coupling $\lambda$ must be split into $k + 1$ different couplings $\lambda_0,\dots,\lambda_k$. This introduces a first (trivial) source of 
explicit breaking and, naively, one could think this is enough to our purposes. So, as a first step, let us suppose that this is the only explicit breaking source. 
If we collect the fields $S'$ and $V'$ in a vector
\be
\Phi^T \equiv \left(S_1,\dots , S_k, V_1, \dots , V_k \right)~,
\ee
and repeat the same analysis of the previous section, we end up at low energy with the following O'R model
\be
W_{eff} = fS_0 + S_0 \sum_{I,J=1}^k C^{IJ}V_I V_J - \sum_{I=1}^k \lambda_I \Lambda S_I V_I
\ee
whose mass and Yukawa matrices are
\be
M = - \Lambda \, \mathrm{diag}(\lambda_1, \dots , \lambda_k) \otimes \left(\begin{array}{cc}
											     0 & 1 \\
											     1 & 0
                 										   \end{array}\right),\quad   N = C \otimes \left(\begin{array}{cc}
											                                                                                  0 & 0 \\
											     								       0 & 1
                 										  								      \end{array}\right)~.
\ee
This corresponds, again, to a theory with $\dt M \neq 0 , \dt N =0$ {\it and} R-charges equal to 0 or 2 only. Hence, less trivial deformations are needed, to reach our goal.

Let us first notice that since $R'(S_0) = R(S_0) = 2 $, whenever a field $V_I$ enters the $C_1$ block, hence appearing quadratically in the Yukawa coupling, then 
$R'(V_I) = R(V_I) = 0$ and $R'(S_I) = R(S_I) = 2$. For such fields, once $G$ and hence $C$ are fixed, there is no possible definition of $U(1)_{R'}$ allowing 
$R'$ charges other than 0 or 2, independently of possible deformations of the superpotential. As a consequence our deformations will focus on the $C_2$ block, only.

Let us suppose there exists an $\boldsymbol{r_I}$ in the $C_2$ block with $\boldsymbol{1} \subset \boldsymbol{r_I}\otimes\boldsymbol{r_I}$ \footnote{If there is no 
irreducible representation in the $C_2$ block with this property, one can even consider reducible ones and proceed along the same lines. For instance when $G = SU(F-1)$, 
as we will discuss later, one takes $\boldsymbol{r_I} = (F-1) \oplus \overline{(F-1)}$. When some $\boldsymbol{r_I}$ is reducible, one should be careful to assign a 
different value of the $\lambda$ coupling to any {\it irreducible} component, in order to ensure genericity.} and let $\boldsymbol{r_J}$ be the representation coupled 
to $\boldsymbol{r_I}$ by $C_2$. We consider two possible modifications of the ITIY theory:
\begin{itemize}
\item[(a)]{Give a (large) mass to the singlet $S_I$ by adding a superpotential term
\be\label{DeltaW}
\Delta W_{tree} = \frac{m_I}{2} {S_I}^2 ~,~ m_I \gtrsim \Lambda ~.
\ee
The effect of this term is to make $R'(S_I) = R'(V_I)=1$, $R'(S_J) = 3$ and $R'(V_J)=-1$. At
low energy $S_I$ can be integrated out and a quadratic term for $V_I$ is generated
\be
\Delta W_{eff} = - \frac{\lambda_I^2 \Lambda^2}{2m_I} V_I^2
\ee
(notice that the mass of $V_I$ is smaller than $\Lambda$). This way, we get at low energy a O'R model having some field with R-charge different from 0 and 2. 
On the other hand, since
the invertibility of the $M$ matrix is not affected by the above deformation, we still have $\dt M \not =0$ (and $\dt N=0$). Hence, deformations of this type give models
with $n=0$.
}
\item[(b)]{Perform deformation (\ref{DeltaW}) for the index $I$ {\it and} eliminate altogether the singlet $S_J$ from the field theory content. One can easily see that 
this modification gives a different theory with $\dt M = 0$, hence models with $n>0$.

Notice that the same effect can obtained sending the coupling $\lambda_J \to 0$. This way, the field 
$S_J$ would remain as a free field, and hence it would not enter the dynamics. However, the theory would loose  
genericity, since there would be no symmetry reasons for the coupling $~ S_J V^J$ to be absent. Dropping the field 
from the theory, instead, while giving rise to the same low energy effective dynamics, keeps the UV theory generic. 
}
\end{itemize}
The correspondence between the above deformations of the UV theory and the type of resulting generalized O'R models one gets at low energy, can be further clarified if 
one considers a configuration of parameters such that all  couples $(S_K,\,V_K)$ which do not undergo any deformation are integrated out. This can be achieved taking 
the corresponding $\lambda_K$ couplings sufficiently large. The resulting theory in terms of the light fields is
\begin{itemize}
\item{type I, if we perform only (a) modifications,}
\item{type II (i.e. $\dt N \neq 0$), if we perform only (b) modifications,}
\item{type III (i.e. $\dt N = 0 $), if we perform both (on different, independent indices).}
\end{itemize}

The first question one should worry about is whether these deformations preserve the supersymmetry breaking mechanism and/or they give rise to supersymmetric vacua. 
Both deformations may allow, in principle, for a non zero VEV for the corresponding mesonic fields, and supersymmetric vacua are restored whenever these VEVs can be
arranged to solve the quantum constraint (\ref{qc}). Indeed, an analysis of the full set of the F-term equations reveals that
\begin{itemize}
\item{Any modification of type (a) introduces a runaway supersymmetric vacuum at $S_0 \to \infty$.}
\item{Any modification of type (b) introduces a runaway supersymmetric vacuum at $S_0 \to 0$.}
\end{itemize}
In fact, a simple argument can be given for the location of runaways along the pseudomoduli space \cite{Ferretti:2007ec}, if one takes into account that $R'(S_0)=2$. In the case (a) the 
runaway is for $V_I \to \infty$ with $R'(V_I) = 1$, and therefore $S_0 \sim V_I V_J^{-1} \to \infty$. In the case (b), instead, it is a field $V_J$ having $R'(V_J)=-1$ 
that goes to infinity, and therefore $S_0 \sim V_IV_J ^{-1} \to 0$.

The upshot is that along the pseudomoduli space there will be a classically stable region together with other regions where instabilities emerge, in agreement with 
the general feature of O'R models with R-charges other than 0 or 2. Instabilities are expected in the region of large pseudomodulus in models with $\dt M \neq 0$ 
and $\dt N=0$, and near the origin if $\dt M = 0$ and 
$\dt N \not =0$. This matches with, respectively, the effect of deformation (a), which does not affect $\dt M$ and gives runaways for $S_0 \to \infty$, 
and deformation (b), which induces $\dt M = 0$ and gives runaways for $S_0 \to 0$. A deformation of type (a) + (b) gives both runaways (though with respect to 
different mesonic directions). 
Notice that the runaways are found in the small $V'$ approximation, hence one has to solve the D-term equations 
along the putative runaway mesonic directions, in order to establish whether these are real runaways in the full theory or they lie at finite distance in field space. 

In concrete examples, one has first to determine which region of the pseudomoduli space is classically 
stable. Then, one should see where (and if) supersymmetry 
breaking vacua are stabilized by quantum corrections, and finally check whether their lifetime is sufficiently long, as well as the 
extent to which K\"ahler corrections coming from gauge dynamics may influence the whole analysis. 

\section{Breaking the flavor symmetry}

In this section we would like to put our strategy at work and consider some concrete examples in detail. We will choose a specific global
symmetry breaking pattern (a group $G$), implement deformations of type (a) and/or (b) compatible with this choice, and look at the low energy
effective theory, once the confining gauge dynamics has taken place. We start by analyzing 
the case where the surviving global symmetry group $G$ is the $SO(F)$ subgroup of $Sp(F)$ specified by the embedding
\be
\label{sof1}
SO(F) \owns O \to \left(\begin{array}{c|c}
                                         O & 0 \\
                                         \hline
                                         0 & O
                                         \end{array}\right) \in Sp(F)~.
\ee
This is possibly the simplest non-trivial choice one can make, but it is rich enough to let us address many of the issues outlined in the previous section.
Moreover, it is a convenient first step for
possible phenomenological applications, since one could easily embed a GUT group into $SO(F)$.
In the second part of this section we will discuss other possibilities for $G$.

\subsection{SO(F) flavor symmetry}

Under the $SO(F)$ defined by the embedding (\ref{sof1}) the $S'$ and $V'$ fields defined in eq.s~(\ref{repvs}) decompose according to
\be
S' = \left(\begin{array}{c|c}
                S_1          & S_3  \\
               \hline
               -{S_3}^T     &  S_2   \\
              \end{array}\right),\,                   V' = \left(\begin{array}{c|c}
                                                                                   V_1     &    V_3 \\
                                                                                   \hline
                                                                                  -V_3^T &   V_2 \\
                                                                                  \end{array}\right)~,
\ee
where $S_1$, $S_2$, $V_1$ and $V_2$ are antisymmetric tensors of $SO(F)$, $S_3$, $V_3$ are traceless tensors, and we have chosen a basis in
which 
\be
J = \left(\begin{array}{c|c}
               0 & \boldsymbol{-1_F} \\
               \hline
               \boldsymbol{1_F} & 0 \\
              \end{array}\right)~.
\ee
The $Sp(F)$ quadratic invariant is rewritten as
\be
\tr\left[J V' J V'\right] = 2\left(V_3^2 - V_1V_2\right)
\ee
where traces on $SO(F)$ indices are understood, so that in the basis $(V_1,\,V_2,\,V_3)$ we have
\be
C = h \left(\begin{array}{ccc}
               0 & -1 & 0 \\
               -1 & 0 & 0 \\
               0 & 0 & 2
               \end{array}\right)~.
\ee
There are now two paths we can follow. This form of the $C$ matrix allows one to consider either a type (a) or type (b) deformation on, say, the index 1 (equivalently the index 2, 
while being $C^{33}\neq 0$, no deformations can be introduced for the index 3). In the former case we obtain the tree level superpotential
\be
W_{tree} = \lambda_{0}\Lambda S_0 V_0 +  \lambda_{1}\Lambda S_1 V_1 +  \lambda_{2}\Lambda S_2 V_2+ \lambda_{3}\Lambda S_3 V_3 + \frac{m_1}{2} {S_1}^2~,
\ee
which is generic under the $SO(F)\times U(1)_{R'}$ global symmetry, with $R'$ charge assignment
\begin{center}
\begin{tabular}[c]{c|cccccccc}
         &$S_0$ & $S_1$ & $S_2$ & $S_3$ & $V_0$ & $V_1$ & $V_2$ & $V_3$  \\
\hline
$R'$ & 2      &  1     &  3     &  2     &   0    &    1   &    -1  &  0
\end{tabular}
\end{center}
As specified in the previous section, we consider $m_1 \gtrsim \Lambda$. Moreover, since $V_3$ is forced to have 0 R-charge and cannot undergo any deformation, 
for simplicity we will take $\lambda_3 \gg \lambda_{1,2} $. This way, we can integrate out $S_1$, $S_3$, $V_3$.

Solving the quantum constraint, at energies below the scale $\Lambda$ one gets the effective superpotential
\be
W_{eff} = f S_0 - 2 h\, S_0 V_1V_2 + \lambda_2\Lambda S_2 V_2 - \frac{\lambda_1^2\Lambda^2}{2m_1} V_1^2~.
\ee
This is an O'R-like superpotential of the general form (\ref{wgenor}), with $S_0$ playing the role of the pseudomodulus. Collecting the other
low energy fields in the vector $\Phi^T \equiv (S_2,\,V_1,\,V_2)$ we get
\be
M = \left(\begin{array}{ccc}
                0 & 0 & \frac{1}{2}\lambda_2 \Lambda  \\
                0 & -\frac{\lambda_1^2\Lambda^2}{2m_1} & 0  \\
                 \frac{1}{2}\lambda_2 \Lambda& 0 & 0
               \end{array}\right)~,~
N = -  h\left(\begin{array}{ccc}
                0      & 0  & 0  \\
                0      & 0  & 1  \\
                0      & 1  & 0
               \end{array}\right)~.
\ee
Hence, we end up with $\dt M\neq0 , \dt N =0$ and R-charges other than 0 or 2, that is a type I model. At the classical level the pseudomodulus is locally 
stable in a finite region around the origin and there is a runaway for $S_0 \to \infty$. A simple version of this superpotential with no flavor symmetry 
was studied in \cite{Shih:2007av}. 

It is perhaps more interesting to choose the other option. If we perform a type (b) deformation on the index 1, we obtain now the following
superpotential
\be
W_{tree} = \lambda_{0}\Lambda S_0 V_0 +  \lambda_{1}\Lambda S_1 V_1 + \lambda_{3}\Lambda S_3 V_3 + \frac{m_1}{2} {S_1}^2.
\ee
Under the same assumptions as before, we are now led to the effective superpotential
\be\label{simple}
W_{eff} = f S_0 - 2 h\, S_0 V_1V_2  - \frac{\lambda_1^2\Lambda^2}{2m_1} V_1^2,
\ee
which is again of the form (\ref{wgenor}), but now we are left with one field less and the matrices $M$ and $N$ take the form
\be
M = \left(\begin{array}{cc}
                 -\frac{\lambda_1^2\Lambda^2}{2m_1}  & 0  \\
                 0  &  0
               \end{array}\right),\,
N = -  h\left(\begin{array}{cc}
                 0  & 1  \\
                1  & 0
               \end{array}\right)~.
\ee
This is a model with $\dt M = 0$ and $\dt N \not= 0$, hence a type II model, with $S_0$ the pseudomodulus. This modified superpotential leads to runaway 
supersymmetric vacua at $V_2 \to \infty$, $S_0 \to 0$ and the pseudomoduli space is classically stable everywhere but in a neighborhood of the origin. An analysis 
of D-equations in terms of the original electric variables reveals that along the D-flat direction $V_2$, an ADS superpotential is generated by the dynamics of 
the unbroken gauge group. Therefore, in this case the approximation of small $V'$ gives a result which is reliable even in the complete theory.

In summary, choosing $G= SO(F)$, we see one can construct models of both type I and II (the symmetry breaking pattern is too simple to allow
for independent deformations of type (a) and (b) so, in order to get type III models one should look for less simple global symmetry breaking patterns).
The question of the actual existence of the local supersymmetry breaking vacua and their lifetime can be addressed with a calculation of the Coleman-Weinberg 
potential, and by evaluating, possibly, the magnitude of uncalculable K\"ahler corrections around such minima. In the following we address these 
two issues in turn.

\subsubsection{Loop corrections and the metastable vacuum}

In what follows we would like to show that the model  $\eqref{simple}$ develops a parametrically long lived, R-symmetry breaking metastable vacuum at one loop.
One should compute the Coleman-Weinberg (CW) potential
\be\label{CW}
V_{CW}=\frac{1}{64\pi^{2}}Tr\left[\mathcal{B}^{2}\log\frac{\mathcal{B}}{\Lambda^{2}}-\mathcal{F}^{2}\log\frac{\mathcal{F}}{\Lambda^{2}}\right]
\ee
along the pseudoflat direction, where $\mathcal{B}$ and $\mathcal{F}$ are the (mass)$^2$ matrices for bosons and fermions, respectively, and depend
on the value of the pseudomodulus $S_0$. Up to inessential overall group factors, our computation resemble a similar one done in
\cite{Cheung:2007es,Carpenter:2008wi}.

As we have already noticed, some of the eigenvalues of the scalar mass matrix become zero at a certain value of the pseudomodulus VEV, $X_{min}$,
meaning that the pseudomoduli space is locally stable only for $S_0>X_{min}$. Below this value the system is classically driven toward the runaway configuration.
Substituting the mass eigenvalues into $\eqref{CW}$ one finds
\be
V_{CW}=\frac{F(F-1)}{2}~\frac{\tilde{m_1}^{4}y^{2}}{32\pi^{2}}\left[2\log\left(\frac{\tilde m_1^{2}}{\Lambda^{2}}\right)+g(z)\right]+\mathcal{O}(y^{4})~,
\ee
where
\be\label{dimensionless}
y \equiv \frac{fh}{\tilde{m}_1^2} ~,~ z  \equiv \frac{h\langle S_0 \rangle}{\tilde{m}_1}
\ee
are the supersymmetry breaking scale and the scalar field VEV in units of $\tilde{m}_1=\lambda_1^2\Lambda^2/2m_1$, the function $g$ is
\be
g(z) = \dfrac{1+12z^{2}}{1+4z^{2}}+4\log z+\dfrac{1+2z^{2}}{(1+4z^{2})^{3/2}}\log\dfrac{1+2z^{2}+\sqrt{1+4z^{2}}}{1+2z^{2}-\sqrt{1+4z^{2}}}~,
\ee
and we have made an expansion in the supersymmetry breaking parameter $y$ (we will comment further on this point below and in the following subsection).

A numerical analysis shows that a minimum for $V_{CW}$ exists at $z \approx0.249+\mathcal{O}(y^{2})$. However, as shown in figure 1, in order for this
to be an actual (local) minimum and not a saddle point one must impose
\be
y\lesssim10^{-3}~,
\ee
which is consistent with the small $y$ approximation.
\begin{figure}
\label{plots}
\begin{center}
\includegraphics[height=0.20\textheight]{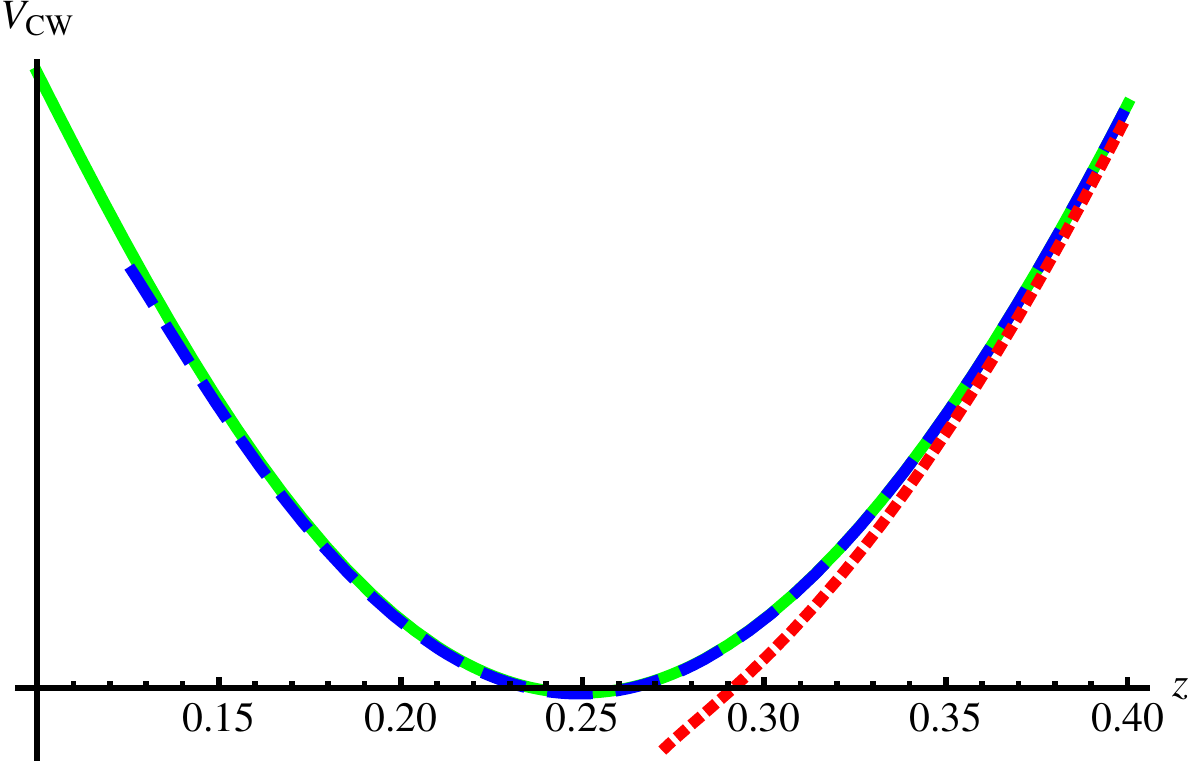}
\caption{\small The effective potential normalized in units of  $\tilde{m}_1^4y^{2}$, for $F = 2$ and $y=10^{-2}$ (dotted line), $10^{-3}$ (dashed line) 
and $10^{-4}$ (solid line). Each
plot ends on the left at the corresponding $z_{min} = z_{min}(y)$. For $y> 10^{-3}$ the would-be minimum would falls into the unstable region of the classical pseudomoduli
space and the theory does not have a metastable vacuum. For smaller values of $y$ the minimum exists, and a potential barrier develops against decay toward the supersymmetric
runaway vacua at $S_0 \to 0$.}
\end{center}
\end{figure}

A rough estimate of the parametric dependence of the lifetime can be given by noticing that, keeping fixed the vacuum energy density $f^2$,
the barrier width scales like
\be
X_{*}-X_{f} \sim\sqrt{\frac{f}{\lambda}}(0.249-y^{\frac{1}{3}})y^{-\frac{1}{2}}\sim y^{-\frac{1}{2}}
\ee
where $X_*$ is the value of the pseudomodulus at the local minimum of the potential, and $X_{f}$ is the value at which the potential
along the runaway direction becomes equal to $f^{2}$, the energy density of the metastable vacuum. This indicates that the lifetime is parametrically long in
the limit of small $y$.

\subsubsection{K\"ahler corrections and calculability}

The ITIY model, as well as any of the deformations we presented in section 3, is an instance of uncalculable DSB model. Therefore, one should worry whether
 K\"ahler  potential corrections coming from gauge theory dynamics at scale $\gtrsim\Lambda$, could affect the low energy effective theory and spoil 
the quantum analysis performed above. Following the discussion of \cite{Chacko:1998si}, we want to estimate such corrections and compare them to the one loop effective 
potential contributions. If the latter are dominant, at least in some region of parameter space, then the calculation performed in terms of the low energy degrees of 
freedom is reliable and the metastable vacuum survives the embedding in the UV theory.

For definiteness, we keep on focusing on the example $\eqref{simple}$, but most of our considerations have wider applicability. Since we are interested in the quantum 
lifting of the tree-level pseudo-flat direction, we can restrict the K\"ahler potential to the pseudomoduli space, after the massive fields $S_1$, $S_3$ and $V_3$ have 
been integrated out. First, notice that the holomorphic decoupling of such fields is expected to produce non-canonicity of the effective K\"ahler potential. However, 
these corrections are largely suppressed in the hierarchical regime
\be
\lambda_{3} \gg \lambda_1 ~, ~ m_1 \gg \lambda_1\Lambda~,
\ee
in which those fields can be integrated out. The K\"ahler potential for the remaining fields is constrained
by the global symmetry to have the form (recall that we have chosen a point of maximal symmetry on the moduli space, which constrains the
K\"ahler potential to be diagonal in the effective fields)
\be
K = {S_0}^{\dagger}S_0 + V_1^{\dagger}V_1+\ V_2^{\dagger}V_2 + \Lambda^{2}~{\cal G}(h S_{0}/\Lambda,h {S_0}^{\dagger}/\Lambda)~,
\ee
where
\begin{itemize}
\item the real function ${\cal G}$ is parametrizing our ignorance of the gauge loop corrections, and depends only on $S_{0}$ since we are restricting to the
pseudomoduli space;
\item the prefactor $\Lambda^{2}$ gives vanishing corrections in the classical limit $\Lambda\to0$;
\item the combination $h S_{0}/\Lambda$ appears because the only way gauge interactions know of the singlet is through
the tree level quark masses $\sim h \langle S_0 \rangle$.
\end{itemize}
This shows that the first corrections are of the form
\be
{\cal G}\sim\frac{h^{4}(S_0{S_0}^{\dagger})^{2}}{\Lambda^{4}}+\mathcal{O}\left(\frac{h^{6}(S_0{S_0}^{\dagger})^{3}}{\Lambda^{6}}\right)~,
\ee
giving a term in the effective potential of order
\be
\Delta V=-\Lambda^{2}\left(\partial_{S_0}\partial_{{S_0}^{\dagger}}{\cal G}\right)|\partial_{S_{0}}W_{eff}|^{2}\sim\Lambda^{2}h^{6}|S_{0}|^{2}~.
\ee
On the other hand, the CW contribution is of order $\sim m^{4}$ where $m$ is the typical mass of the light IR degrees of freedom entering the loops. In the present 
case, these masses are given by $\tilde{m}_1= \lambda_1^2\Lambda^2/2m_1$ and $h \langle S_{0} \rangle$. Therefore,
suppression of uncalculable corrections requires
\be
 h \Lambda   \ll \langle S_0 \rangle  \ll   \frac{\tilde{m}_1^2}{h^3 \Lambda } = \frac{\lambda_1^4 \Lambda^{3}}{4h^3m_1^2}
\ee
which, in terms of dimensionless parameters $\eqref{dimensionless}$, recalling the definitions (\ref{fh}), reads
\be
h\sqrt{y/4F} \ll z  \ll  \frac{1}{h \sqrt{y/4F}}~.
\ee
Since the local minimum seats at $z_{min}\simeq0.249 + \mathcal{O}(y^{2})$ these inequalities are trivially satisfied in the limit of small $y$ and $h$. It is 
amusing to notice that the limit of small $y$ provides both a long lifetime and small K\"ahler corrections.

Let us recap the discussion above and take a closer look to the hierarchies we need, to let the theory having a safe 
local minimum. First, when giving a mass to the singlet $S_1$, we have chosen $m_1\gtrsim \Lambda$. This ensures that at
low energy $S_1$ can be integrated out and the corresponding mesonic field $V_1$ gets a small mass $\tilde{m}_1 = \lambda_1 \Lambda^2/2m_1$ (we will always consider
$\lambda_I < 1$ so that all dynamically generated masses $\lambda_I \Lambda$ are below the dynamical scale). Then, for simplicity, we have chosen the field pair
which is not modified, ($S_3,V_3$), to be much heavier than the other pairs, and this can be accomplished by a larger value for the corresponding coupling,
$\lambda_3 \gg \lambda_{1,2}$. As we have seen, the existence of a
local minimum with a long lifetime and suppressed uncalculable corrections are both
controlled by the smallness of one single parameter,
\be
y = \frac{f h}{\tilde{m}_1^2} = 2 \left(\frac{\lambda_0}{\lambda_1^2}\right)^2\left(\frac{m_1}{\Lambda}\right)^2~.
\ee
The requirement of small $y$ forces a small value for $\sqrt{\lambda_0}$ : for instance, if $m_1/\Lambda \sim 10$ then $y \lesssim 10 ^{-3}$ implies
$\sqrt{\lambda_0}\lesssim 10^{-1.25} \lambda_1$. Notice that as for any dynamical model, in this model both the supersymmetry breaking scale
$f = \sqrt{2F}\lambda_0 \Lambda^2$ and the masses of the low energy O'R model $\lambda_I \Lambda $ are related to one and the same dynamical scale $\Lambda$.
Therefore, it is not surprising that a (modest) tuning between dimensionless
parameters is necessary to obtain metastable supersymmetry breaking. The limit of small $y$ can indeed be simply reinterpreted as the limit of small vacuum energy
with respect to the scale set by the masses, and, from the expressions of $f$ itself, it is clear that this requires $\sqrt{\lambda_0}$ to be small compared to all
other $\lambda$'s.

\subsection{Other breaking patterns}

While the class of models we discussed above is general enough to make our strategy manifest, it is clearly not the most general option one can think of.
For instance, as we have already noticed, the possibility of making independent deformations of type (a) and (b) requires a more involved symmetry breaking pattern.
Moreover, in view of phenomenological applications,  having $SU$ global symmetry groups, besides $SO$ groups, might also be interesting. In what follows, we 
want to make a few comments on both these options. We will not discuss the vacuum structure in any detail, nor the dynamics around the supersymmetry breaking minima, 
but just limit ourselves to 
display the basic structure of the emerging low energy effective theories.

\subsubsection*{$\bullet \;SO(n) \times SO(F-n) $}
The simplest step we can take beyond the $SO(F)$ models we analyzed before, is to consider the group $G$ to be $G = SO(n) \times SO(F-n) $, with $1 < n < F$. 
Under such $G$, the $SO(F)$ components of $V$ and $S$ defined in eqs.(\ref{repvs}) decompose as follows
\be
V_I = \left(\begin{array}{c|c}
                  V_I^{(n)} & W_I \\
                  \hline
                  -W_I^T & V_I^{(F-n)}
                  \end{array}\right)\,,\quad
S_I = \left(\begin{array}{c|c}
                  {S_I}^{(n)} & T_I \\
                  \hline
                  -T_I^T & {S_I}^{(F-n)}
                  \end{array}\right)\,,\quad I = 1,\,2,\,3~,
\ee
where $W_I$ and $T_I$ are $n\times (F-n)$ ``bi-vectors" of the two $SO$ factors. In the basis
\be
(V_1^{(n)},\, V_2^{(n)},\, V_1^{(F-n)},\, V_2^{(F-n)},\, W_1,\, W_2)
\ee
the matrix $C_2$ takes the form
\be
C_2 = h \left(\begin{array}{cc|cc|cc}
               0  & -1 &  &  &  &  \\
               -1 & 0  &  &  &  &  \\
               \hline
                &   & 0  & -1 &  &  \\
                &   & -1  & 0 &  &  \\
               \hline
               &  &  &  & 0 & 1 \\
               &  &  &  & 1 & 0
               \end{array}\right)~.
\ee
Clearly, there are now much more options for the modified theory: each off-diagonal two-by-two block can undergo deformations (a) or (b). Let us consider a
possibility which was not available in the simpler case $G=SO(F)$, and make a deformation (a) for the first two blocks together with a deformation (b) for
the last one. The starting point is therefore a generic tree level superpotential of the form
\bea
W_{tree} = \lambda_0 \Lambda S_0 V_0 + \Lambda \sum_{I=1}^3\left[\lambda_I^{(n)} V_I^{(n)} {S_I}^{(n)}+ \lambda_I^{(F-n)} V_I^{(F-n)} {S_I}^{(F-n)} \right] \nonumber \\
+ \lambda'_1\Lambda T_1 W_1 + \lambda'_3\Lambda T_3 W_3 +  \frac{m_1^{(n)}}{2} {{S_1}^{(n)}}^2 + \frac{m_1^{(F-n)}}{2} {{S_1}^{(F-n)}}^2 + \frac{m'_1}{2} T_1^2~.
\eea
Just like in the simpler $SO(F)$ case, no deformations can be made for the $I = 3$ fields and, for simplicity, we choose the parameters so that these fields can be 
integrated out.
Solving the quantum constraint and integrating out all other heavy fields, one finally gets the effective superpotential
\bea
W_{eff} = f S_0 + 2h S_0 \left[ W_1 W_2 - V_1^{(n)}V_2^{(n)} - V_1^{(F-n)}V_2^{(F-n)} \right] \nonumber \\
 + \lambda_2^{(n)} \Lambda {S_2}^{(n)} V_2^{(n)}  + \lambda_2^{(F - n)} \Lambda {S_2}^{(F - n)} V_2^{(F - n)} \nonumber \\
 - \frac{{\lambda_1^{(n)}}^2\Lambda^2}{2m_1^{(n)}} {V_1^{(n)}}^2 - \frac{{\lambda_1^{(F-n)}}^2\Lambda^2}{2m_1^{(F-n)}} {V_1^{(F-n)}}^2 - 
\frac{{\lambda'}^2\Lambda^2}{2m'} W_1^2~,
\eea
which describes a model with $\dt M, \det N=0$, i.e. a type III model, in the terminology of \cite{Cheung:2007es} . Notice, in passing, that the 
number of light fields typically diminishes, the more deformations (a) and/or (b) one does. This might be a welcome feature for phenomenological applications.

\subsubsection*{$\bullet \; SU(F-1)$}

Here we consider a particular breaking pattern leading to theories with an $SU(F-1)$ flavor symmetry. First we introduce the embedding
in $Sp(F)$
\begin{equation}
SU(F-1) \owns U\to\left(\begin{array}{ccc}
\boldsymbol{1_{2}} & 0 & 0\\
0 & U & 0\\
0 & 0 & U^{*}
\end{array}\right)~.
\end{equation}
The field content can be arranged in self-conjugate (reducible) representations in the following way
\bea
\begin{array}{c|c}
           & SU(F-1)\\
\hline
V_{1} & \bullet\\

V_{2} & \mathbf{\left(F-1\,\oplus\,\overline{F-1}\right)\otimes_{A}\left(F-1\,\oplus\,\overline{F-1}\right)}\\

V_{3} & \mathbf{F-1\,\oplus\,\overline{F-1}}\\

V_{4} & \mathbf{F-1\,\oplus\,\overline{F-1}}\\

S_{1} & \bullet\\

S_{2} & \mathbf{\left(F-1\,\oplus\,\overline{F-1}\right)\otimes_{A}\left(F-1\,\oplus\,\overline{F-1}\right)}\\

S_{3} & \mathbf{F-1\,\oplus\,\overline{F-1}}\\

S_{4} & \mathbf{F-1\,\oplus\,\overline{F-1}}
\end{array}
\eea
The ITIY superpotential in terms of these fields reads
\begin{equation}
W_{ITIY}=\lambda_{0}\, S_{0}V_{0}+\sum_{I=1}^4 \lambda_{I}\, S_{I} V_{I}~.
\end{equation}
Before introducing deformations this superpotential still has a full $Sp(F-1)\times SU(2)$ global symmetry \footnote{The $SU(2)$ factor rotates the indices $I=3,4$. 
In particular, they have $+$ respectively $-$ charge under the diagonal generator
$\sigma_{3}$. This is the charge we are going to mix with the $R$.} where the $Sp(F-1)$ invariant product is specified as follows: for $I=1$ it is the trivial 
multiplication, for $I=2$ it corresponds to ${\rm tr}_{2(F-1)}\left[J_{2(F-1)}S_{2}J_{(F-1)}V_{2}\right]$, and for $I=3,4$ to ${S_I}^{T}J_{2(F-1)}V_{I}$. The 
quadratic invariant is now
\begin{equation}
{\rm tr_{2F}}\left[J_{2F}V'J_{2F}V'\right]=2V_{1}^{2}+V_{2}^2+2V_{4}V_{3}-2V_{3}V_{4}=2V_{1}^{2}+V_{2}^2+4V_{4} V_{3}~,
\end{equation}
and from it we can read the matrix $C$
\begin{equation}
C=h\left(\begin{array}{cccc}
2 & 0\\
0 & 1\\
 &  & 0 & 2\\
 &  & -2 & 0
\end{array}\right)~.
\end{equation}
The only possible deformation involves the fields $3,4$ and it can be of type (a) or (b). This deformation breaks $Sp(F-1)\times SU(2)\to SU(F-1)$ and forces the R-charge
of $V_{3(4)}$ to be $1(-1)$, whereas the other $V$ fields remain uncharged.

In order to implement the deformation, we write in this case a quadratic term for $S_{3}$ which breaks $Sp(F-1)$ while preserving a $SU(F-1)$. The choice is 
unique (up to a multiplicative factor) and reads
\begin{equation}
S_3^2 \equiv {S_3}^{T}\left(\begin{array}{cc}
0 & \boldsymbol{1_{F - 1}}\\
\boldsymbol{1_{F - 1}} & 0
\end{array}\right)S_{3}~.
\end{equation}
Adding a large mass term of this form to the superpotential we obtain\footnote{This superpotential is not generic since there are allowed couplings of the form $S_0V_1$ 
and $S_1V_0$ which are not present. Nevertheless, once we integrate out $V_1, S_1$, the only effect of these extra terms is a redefinition of the coupling $\lambda_0$.}
\begin{equation}
W_{tree}=\lambda_{0}\, S_{0}V_{0}+\lambda_{I}\, S_{I} V_{I}+\frac{m}{2}S_{3}^2.
\end{equation}
After integrating out $S_{3},S_{1},S_{2},V_{1},V_{2}$ we end up with a type I O'R superpotential
\begin{equation}
W_{eff}=f\, S_{0}+4h\, S_{0}V_{4} V_{3}+\Lambda\lambda_{4}\, S_{4} V_{4}+\frac{\Lambda^{2}\lambda_{3}^{2}}{m}V_{\text{3}}^2~,
\end{equation}
or a type II superpotential
\begin{equation}
W_{eff}=f\, S_{0}+4h\, S_{0}V_{4} V_{3}+\frac{\Lambda^{2}\lambda_{3}^{2}}{m}V_{\text{3}}^2~,
\end{equation}
if we perform a (b) deformation, instead. Restoring the irreducible field content the last equation reads
\begin{equation}
W_{eff}=f\, S_{0}+4h\, S_{0}\left(\tilde{V_{4}}V_{3}-\tilde{V_{3}}V_{4}\right)+2\frac{\Lambda^{2}\lambda_{3}^{2}}{m}\tilde{V_{\text{3}}}V_{3}~,
\end{equation}
where tilded fields transform in the anti-fundamental and untilded in the fundamental of $SU(F-1)$. This looks like the case $r=n=2$ discussed in
\cite{Cheung:2007es}.

\section{Conclusions}

In this paper we have constructed DSB models which at low energy reduce to WZ models admitting supersymmetry breaking vacua where also the R-symmetry is spontaneously 
broken. Starting from well known generalizations of the ITIY  model, we have proposed a precise pattern to modify the microscopic theory so to get at low energy 
models falling in all three classes of extraordinary gauge mediation \cite{Cheung:2007es}. In the second part of the paper we focused on a concrete (class of) 
example(s) and discussed in some detail the low energy effective theory both at classical 
and quantum level. Interestingly, the same window which allows for a long-lived metastable vacuum, makes the perturbative analysis of the effective theory reliable 
against corrections coming from $\Lambda$-dependent K\"ahler potential contributions.

Our results can be generalized in various ways. For one thing, one could consider ITIY models with  $SO(N)$ or $SU(N)$ gauge groups, and see whether a general pattern 
like the one spell-out in section 3, still emerges. In principle, there's nothing special about $Sp(N)$. A second, 
more interesting thing to do would be to consider more general global symmetry breaking patterns, like those we commented upon in section 4.3. As we have seen, one could break 
the global $Sp(F)$ symmetry down to  $SO(F-n) \times SO(n)$ or, more generally, to several $SO$ factors. Or consider unitary global symmetry groups. This opens-up more 
possibilities to modify the tree level superpotential of the UV theory, and let one cover a larger class of hidden sector effective models, including genuine type 
III models. 

On a more phenomenological side, our models are promising as direct gauge mediation models. Of course, when it comes to construct fully 
fledged phenomenological models, one should take care of many issues. For instance, in models of direct mediation a thing one should worry about 
are Landau  poles. We do not see any major obstacle to envisage a breaking pattern with e.g. $G=SO(F-n) \times SO(n)$, weakly gauge
the $SO(n)$ GUT group (take for definiteness $n=10$) and implement enough deformations of type ({\it a}) and/or ({\it b}) so not to have too many messengers 
around. Seemingly, one could consider to break the original global symmetry group to several unitary groups, e.g. $G=SU(F-1-n) \times SU(n)$ (take now $n=5$). 
Notice that since the UV theory is non-chiral, all messengers would come into real representations so one should not worry about $SU(5)$ gauge anomalies.

Our proposal seems flexible enough to let one cover a sizeable region of the parameter space of supersymmetry breaking models admitting a weakly coupled low 
energy description, whose phenomenology was discussed in \cite{Cheung:2007es}. On the other hand, our models are more constrained since low energy parameters 
depend on the UV physics, and only a more detailed analysis could tell how large this region can be. Here, we just want to emphasize one basic 
phenomenological potential outcome of the results of this work, which is to provide, in principle, complete models of direct gauge mediation with spontaneous 
R-symmetry breaking and unsuppressed gaugino mass.

\section*{Acknowledgments}

We thank David Shih for bringing our attention to this problem, and Zohar Komargodski for enlightening discussions. We also thank 
Marco Serone for useful comments. Finally, we are grateful to Riccardo Argurio, Zohar Komargodski and David Shih for a critical reading of the draft.


\appendix


\end{document}